\documentclass[conference, anonymous]{IEEEtran}

\usepackage[letterpaper,top=2cm,bottom=2cm,left=3cm,right=3cm,marginparwidth=1.75cm]{geometry}

\usepackage[utf8]{inputenc}
\usepackage{array}
\usepackage{booktabs}
\usepackage{multicol}
\usepackage{multirow}
\usepackage{tabularx}
\usepackage{longtable}
\usepackage{threeparttable}
\usepackage{enumitem} 
\usepackage[many]{tcolorbox}
\usepackage{balance}
\usepackage{hyperref}
\usepackage{wrapfig}

\newtcolorbox{fancybox}[1][]{
  enhanced,
  attach boxed title to top center={yshift=-3mm,yshifttext=-1mm},
  colback=green!5!white,
  colframe=green!25!black,
  colbacktitle=green!25!black,
  fonttitle=\bfseries,
  title=#1,
  boxed title style={size=small, colframe=green!45!black, colback=green!45!black, center},
  drop fuzzy shadow,
  width=0.5\textwidth,
  breakable,
  left=0mm,  
  fontupper=\fontsize{9.1}{11}\selectfont  
}

\newcommand{\sectopic}[1]{\vspace{0.2em}\par\noindent{\textit{\bfseries #1}}}

\newcommand{\appName}{\textit{LIEF\_POST}}
\newcommand{\approach}{\textit{RAGTAG}}

\newcommand{\projOne}
{\textit{$Proj_{A}$}}
\newcommand{\projTwo}{\textit{$Proj_{B}$}}

\newcommand{\company}{{Austrian Post}}

\begin{document}
\pagenumbering{arabic}
%
\title{
Generating Test Scenarios from NL Requirements using Retrieval-Augmented LLMs: An Industrial Study}


\author{
    \IEEEauthorblockN{
    Chetan Arora\IEEEauthorrefmark{1}, 
    Tomas Herda\IEEEauthorrefmark{2}, 
    Verena Homm\IEEEauthorrefmark{2}
    }
    \IEEEauthorblockA{\IEEEauthorrefmark{1}Monash University, Melbourne, Australia}
    \IEEEauthorblockA{\IEEEauthorrefmark{2}Austrian Post Group IT, Vienna, Austria}
    Email: chetan.arora@monash.edu, tomas.herda@post.at, verena.homm@post.at
\vspace*{-1em}
}


%


\maketitle

\thispagestyle{plain}
\pagestyle{plain}

\begingroup\renewcommand\thefootnote{\textsection}
\endgroup

\begin{abstract}
Test scenarios are specific instances of test cases that describe a sequence of actions to validate a particular software functionality. 
By outlining the conditions under which the software operates and the expected outcomes, test scenarios ensure that the software functionality is tested in an integrated manner. 
Test scenarios are crucial for systematically testing an application under various conditions, including edge cases, to identify potential issues and guarantee overall performance and reliability. Manually specifying test scenarios is tedious and requires a deep understanding of software functionality and the underlying domain. It further demands substantial effort and investment from already time- and budget-constrained requirements engineers and testing teams. This paper presents an automated approach (\approach)~for test scenario generation using Retrieval-Augmented Generation (RAG) with Large Language Models (LLMs). RAG allows the integration of specific domain knowledge with LLMs' generation capabilities. We evaluate \approach~on two industrial projects from Austrian Post with bilingual requirements in German and English.  
Our results from an interview survey conducted with four experts on five dimensions -- relevance, coverage, correctness, coherence and feasibility, affirm the potential of \approach~in automating test scenario generation. Specifically, our results indicate that, despite the difficult task of analyzing bilingual requirements, \approach~is able to produce scenarios that are well-aligned with the underlying requirements and provide coverage of different aspects of the intended functionality. The generated scenarios are easily understandable to experts and feasible for testing in the project environment. The overall correctness is deemed satisfactory; however, gaps in capturing exact action sequences and domain nuances remain, underscoring the need for domain expertise when applying LLMs. 

\end{abstract}


%
\IEEEpeerreviewmaketitle

\section{Introduction}~\label{sec:introduction}
Test scenario generation, a critical phase in the software quality assurance process, involves specifying situations and corresponding action sequences to validate a software application's functionality and performance against its requirements~\cite{desikan2006software,crispin2009agile}. The essence of test scenarios lies in their ability to simulate diverse user interactions and system behaviors, ensuring the software's reliability in real-world conditions~\cite{desikan2006software,unterkalmsteiner2014taxonomy}. Furthermore, test scenarios allow developers to identify and address potential issues early in the development cycle, significantly reducing software maintenance costs and time. By offering means for evaluating critical functionalities and user interactions, test scenarios enhance the overall quality of the software product, contributing to a more satisfactory end-user experience.

Based on project requirements and the development environment, the test scenarios (\emph{TSs}) can be specified in different formats, e.g., sequence of steps in a UML activity or sequence diagrams~\cite{gordon2009generating,nayak2011synthesis}, formal notations~\cite{cunning1999test,klischat2019generating}, other structured or templated notations~\cite{bertolino2006product,wang2020automatic}, or a combination of aforementioned formats supported with natural language (NL) annotations~\cite{hois2014natural} and scenarios specified in NL~\cite{cem2013introduction}. 

Irrespective of the specification format, manually creating \emph{TSs} is an intellectually intensive task, requiring deep domain and project knowledge, and a significant investment of time and resources. Manually specifying \emph{TSs} has further challenges, such as the potential for overlooking edge cases, the difficulty in maintaining scenarios as requirements change, and the scalability of the process for large projects. An automated approach for generating \emph{TSs} from requirements is thus required to address these challenges, cover all functional aspects, and identify and include edge cases that manual processes might miss to bridge the gap between the functional expectations outlined in the requirements and their practical validation through testing. Such automated solutions can help develop robust and reliable software products capable of performing under a wide range of conditions, thereby increasing user trust and satisfaction.

This industry-innovation paper investigates the automated generation of \emph{TSs} from NL requirements using advanced natural language processing (NLP) techniques, specifically Large Language Models (LLMs). LLMs, trained on vast text corpora, can understand and interpret complex software requirements~\cite{Arora:23}, translating them into detailed, nuanced \emph{TSs}. While LLMs excel in these aspects, they sometimes struggle with domain-specific queries and generate inaccurate information~\cite{kandpal2023large}. A widely-known issue with generative LLMs is the generation of incorrect information, or ``hallucinations'', especially when the generation task at hand extends beyond the LLMs' training data~\cite{lewis2020retrieval}, e.g., the \emph{TS} generation task for a specific domain. For addressing this, we leverage Retrieval-Augmented Generation (RAG)~\cite{lewis2020retrieval} to enhance LLMs. RAG integrates external data retrieval into the text generation process of LLMs, enhancing their ability to generate accurate and relevant responses~\cite{gao2023retrieval}. This combination of RAG and LLMs allows the model to pull relevant information from a curated database, mitigating issues related to incorrect information and ensuring the generation of accurate and domain-specific TSs.

This applied research work has been developed and conducted in close collaboration with the Austrian Post Group IT, the IT division of the larger Austrian Post organization. Below, we provide an insight into the industrial context of this research (Section~\ref{subsec:Industry_Context}) and also cover a working example for this paper (Section~\ref{subsec:Example}).

\begin{figure}[!t]
  
\begin{fancybox}[Example Requirements]

\textbf{R1:} The \appName~shall enable switching the Umsatzsteuer (USt.-Satz) from 20\% to 18\% for the Zollauschlussgebiet towns, listed in Austrian Post documentation in the ZAB list.

\vspace*{0.5em}
\textbf{R2:} The \appName~shall only display the Switch-button in Zustellbasen where the backend flag ``Änderung auf 18\% USt'' is enabled.

\vspace*{0.5em}
\textbf{R3:} The \appName~shall allow switching the USt for both R\"{u}cksendung Inland and Ausland.

\end{fancybox}

\begin{fancybox}[Example Test Scenario]
\textbf{TS1:} Delivery with R\"{u}cksendung Ausland.

\begin{enumerate}[leftmargin=*]
    \item Open the \appName~app in VB mode.
    \item In VB mode, scan the packet with R\"{u}cksendung Ausland flag.
    \item Set the Zustellbasen to a town in the ZAB list.
    \item Long press on the packet and open the details.
    \item Turn the Switch to ON for the 18\% USt.
    \item Verify that the USt rate is displayed as 18\% next to the switch.
    \item Verify that the 20\% USt rates are correctly converted to 18\% USt rates when the Switch is turned ON.
    \item Move the \appName~to ZL mode.
    \item Repeat the test scenario for other towns in the ZAB list to ensure compatibility.    
\end{enumerate}
\end{fancybox}
\caption{Example Requirements and Test Scenario from \projOne~in \appName~(Details in Section~\ref{subsec:Example}).}
  \label{fig:example}
\end{figure}
\subsection{Industry Context}~\label{subsec:Industry_Context}
Austrian Post is an international postal, logistics, and service provider, and Group IT is the division that manages most of the company's IT operations. At Austrian Post, \appName~is a flagship software application (anonymized for data confidentiality reasons) used for all delivery and postage processes. \appName~is available for a special hand-held device used by the deliverers. Given the large scale of \appName, new functionality is regularly added to the software application while maintaining the previous functionalities. Each such substantial functionality addition in \appName~is termed as a \emph{project}, as multiple software engineering (SE) teams with requirements engineers, testers, developers and project and test managers are dedicated to a project. 
Two such projects are the subject of evaluation in this paper -- referred to as \projOne~and \projTwo~in the paper. \projOne~refers to a project in \appName~related to the application and recalculation of value-added tax (VAT or Umsatzsteuer / USt. in German) for specific regions with special VAT arrangements. \projTwo~deals with automated pre-sorting of posts before delivery
for an efficient delivery process.
The requirements for most projects in \appName, including \projOne~and \projTwo, are specified bilingually - a mix of German and English. 
The rationale behind the bilingual requirements is the historical context of operations at Austrian Post (in German) and the internationalization of the organization in terms of the team structure (in English). In addition, some projects are outsourced to external contractors, wherein the interactions are often in English.
However, most \emph{TSs} are specified in English, with the project's key vocabulary (e.g., system names and project keywords) maintained primarily in German. Such bilingual requirements are commonplace at Austrian Post and many other organizations, adding to the complexity of building automated solutions for RE processes.

\subsection{Working Example}~\label{subsec:Example}
Fig.~\ref{fig:example} shows example requirements and an example \emph{TS}, closely adapted (and anonymized while maintaining the intent and structure) from the original requirements and test scenarios in \projOne. The requirements refer to features for applying and recalculating VAT for specific regions. We have preserved the bilingual nature of the requirements for exemplification. R1 refers to enabling VAT calculation switch in special regions, listed in a list (\textit{Zollauschlussgebiete} in German or \textit{ZAB}) from Austrian Post. R2 refers to displaying the switch only for delivery bases (\textit{Zustellbasen} in German) where the change of VAT (18\%) is enabled. R3 focuses on return for both in-land and out-of-the-country deliveries. The example \emph{TS} lists the steps for testing the features listed in the requirements for out-of-the-country return delivery cases. \textcolor{black}{We note that \emph{TSs} typically cover the actions required and the expected results. Some \emph{TSs} have separate expected results and actions subsections formatted, while others combine the actions and expected results as in the example. In our projects, we had both kinds and we did not alter any \emph{TSs} to maintain the realistic settings of our experiment.}

As evident in Fig.~\ref{fig:example}, one of the main challenges in analyzing these requirements is bilingual requirements, which are commonplace for Austrian Post and several other organizations across the globe due to the working context and globalization efforts. Another noteworthy challenge in building an automated solution for generating \emph{TSs} from these real requirements is that the requirements are intricately entwined with domain-specific knowledge, often beyond the information readily available in public sources. For instance, while some domain information related to the postage and delivery is publicly available, most internal processes at Austrian Post or \appName~remain proprietary. This also means that the pre-trained LLMs have a limited understanding of these processes. For example, in the example \emph{TS} in Fig.~\ref{fig:example}, keywords and processes are specific to \appName. \textit{VB} mode (short for \textit{Vorbereitung} in German), in step 1 of the \emph{TS}, refers to the preparation phase in \appName. Similarly, \textit{R\"{u}cksendung Ausland} flag, in step 2 of the \emph{TS}, refers to the fact that the package deals with an out-of-the-country return delivery case. The \textit{Switch}, in step 5 of the \emph{TS}, refers to a specific button and corresponding flag in the backend in \projOne. In addition, the domain understanding that the example \emph{TS} is only complete if the \appName~app is switched from \textit{VB} to \textit{ZL} mode (short for \textit{Zustellung} in German), i.e., the delivery mode, is missing in pre-trained LLMs and the example requirements. For all this information, one would require access to additional domain information to be able to generate accurate and complete \emph{TSs}. 
To this end, we opted for an LLMs and RAG-based solution for investigating this problem, as compared to other possible NLP solutions~\cite{sabetzadeh2024practical}, because (i)~this is a generation task best suited for LLMs; (ii)~the requirements are written in NL, which can be ambiguous and open to multiple interpretations. The issue is further exacerbated in the real-world context of bilingual requirements, where a single requirement mixes two languages - hence, the underlying NLP technique needs to be adept at switching linguistic contexts seamlessly; (iii)~Austrian Post - like several other global organizations - is keen at leveraging (generative) AI techniques to improve the efficacy of the internal processes and overall productivity~\cite{nguyen2023generative}; and (iv) RAG provides an opportunity to integrate the domain knowledge.

\subsection{Contributions}~\label{subsec:Contributions}
\begin{itemize}[leftmargin=*]
    \item We propose a novel approach (\approach) based on large language models (LLMs) and retrieval augmented generation (RAG) for the automated generation of test scenarios from NL requirements. RAG allows leveraging domain-specific information from organizational documents, and pre-trained LLMs help generate test scenarios. 
    \item We report on the design and execution of \approach~in an industrial context, with an evaluation on two real-world projects in the logistics domain with bilingual requirements. We experiment with eight different configurations of \approach, and report on the best configuration. We further report on an interview survey with four experts. Our results show that the experts find the test scenarios generated by \approach, as largely complete in terms of their coverage of all relevant concepts, easy to understand and feasible for execution after having manually addressed some issues related to the correctness of some intermediate steps in the test scenarios. The experts further report that \approach~is helpful in improving their efficiency in specifying test scenarios, compared to manual efforts and also helps highlight quality issues in requirements. 
\end{itemize}

\sectopic{Structure.} Section~\ref{sec:background} provides the background. Section~\ref{sec:approach} discusses our approach (\approach). Section~\ref{sec:evaluation} reports on the design and execution of our industry evaluation on two projects. Section~\ref{sec:threats} discusses threats to validity. Section~\ref{sec:conclusion} concludes the paper. 

\section{Background}~\label{sec:background}
In this section, we provide the background on different concepts in our \emph{TS} generation approach, i.e., large language models (LLMs), prompting, and retrieval augmented generation (RAG). We also briefly cover the related work on this topic in this section.

\subsection{Large Language Models (LLMs)}~\label{subsec:LLMs}

A language model (LM) is a statistical model trained to predict the next word in a sequence and applied for several NLP tasks, i.e., classification and generation~\cite{Jurafsky:20}. For example, given a text sequence, ``Paris is the capital of'', the LMs will predict the next word as ``France'' with the maximum probability. 
LLMs are extensions of LMs trained on a substantially large dataset, with a larger number of weights and parameters, and a complex architecture that can perform various NLP tasks, including text generation, classification and question answering~\cite{brown2020language}. 

\subsection{LLMs Prompt Techniques}~\label{subsec:prompting}
Conversational LLMs (e.g., ChatGPT) need to be provided with instructions or discussion topics by a user in the form of a \emph{Prompt}. The prompts can be designed and refined to achieve the desired outcomes effectively from LLMs. Below, we present some commonly used prompting techniques that are used or evaluated in this paper.

\sectopic{Zero-Shot Prompting} refers to a scenario where an LLM is prompted for a task or a query without any prior examples. The model must rely solely on pre-trained knowledge and understanding to generate a response or solution.

\sectopic{Few-shot prompting} involves providing the LLM with a few examples before prompting it to perform a task. These examples serve as a guide, helping the model understand the expected response or output. It is particularly useful when the LLM must be adapted to a specific type of task or format that it might not have been explicitly trained on, e.g., \emph{TS} generation.



\subsection{Retrieval Augmented Generation (RAG)}~\label{subsec:RAG}
\begin{figure}
  \includegraphics[width=0.46\textwidth]{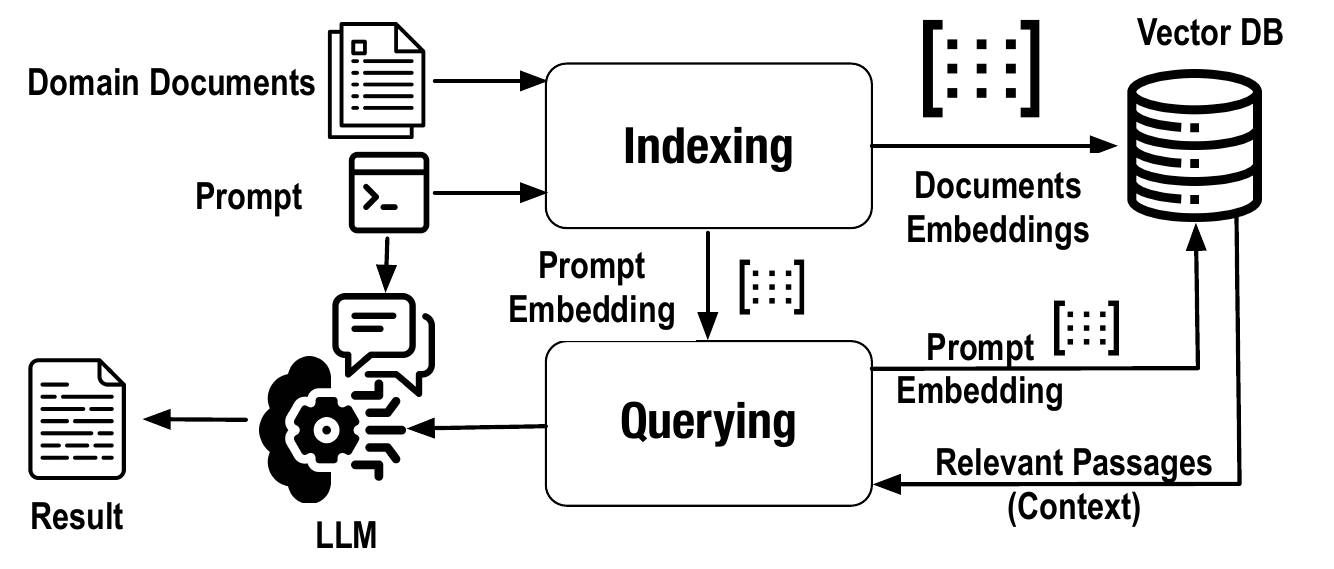}
  \centering
  \caption{Retrieval Augmented Generation (RAG) Pipeline.}
  \label{fig:RAGModel}
\end{figure}

Retrieval-Augmented Generation (RAG) is a hybrid approach that combines the capabilities of two major NLP techniques: information retrieval (IR) and generative LLMs. Generative LLMs like GPT-x models~\cite{brown2020language} have revolutionized NLP domain. LLMs trained on vast datasets generate coherent and contextually relevant text. However, as discussed in Section~\ref{sec:introduction}, LLMs' ability to generate accurate and specific information is often constrained by their training data scope and inherent limitations in accessing external, up-to-date domain information. Traditional IR systems, designed to search and retrieve information from large databases, provide the ability to access specific information. However, they lack LLMs' NL understanding and generation capabilities. 
RAG address these limitations by combining the generative capabilities of LLMs with IR efficiency. When a query is presented to a RAG model, the retrieval system fetches relevant documents or snippets. The retrieved snippets are fed into the generative model, which integrates them into its response generation process. This allows the model to produce responses that are not only contextually rich but also factually accurate. Next, we provide details on the RAG pipeline, visualized in Fig.~\ref{fig:RAGModel}.

\sectopic{Indexing.}
In the RAG pipeline, the indexation process begins with the domain documentation loading, cleansing, and extracting the content. The documents can be loaded in different file formats, e.g., PDF, HTML, and Microsoft Word, and converted into standardized plain text. Thereafter, the text is split into smaller and more manageable passages or chunks (e.g., 512 tokens) to fit within the context limits of different LMs. These passages are subsequently transformed into vector representations through an embedding model, e.g., Sentence Bert (SBERT)~\cite{reimers2019sentence}. These embeddings are numerical representations that encapsulate the semantic content of the passage. An index is created to store the text passages and their vector embeddings as key-value pairs, which allows for retrieving contextually relevant data efficiently and accurately. These key-value pairs can be stored in a vector database to avoid re-indexing the documentation each time. A prompt for the RAG model is indexed using the same embedding model and passed onto the next step in the pipeline.

\sectopic{Querying.} The querying step uses the vector embedding of the prompt to identify the most relevant passages from the vector database with domain documentation embeddings. The similarity is calculated between the two embedding vectors, i.e., the prompt embedding and the 
vectorized passages within the indexed vector database using cosine similarity~\cite{manning1999foundations}. The querying step then retrieves the top-$k$ passages with the highest similarity to the prompt. These passages form the contextual basis for generating the text for the query or the task specified in the prompt.

\sectopic{Generation.} The input prompt and the selected passages from the querying stage are then combined to be passed onto the LLM for response generation. Depending on the task or the query specified in the input prompt, a response could be generated based on the pre-trained model information along with the context passages or from the context passages alone. 

\subsection{Related Work}~\label{sec:related}
Generating test scenarios or testing artefacts from requirements is a relatively less-explored area in RE or SE literature~\cite{mustafa2021automated}. Mustafa et al.~\cite{mustafa2021automated}'s SLR provides an overview of limited automated approaches for test generation from requirements. They identified 30 primary studies, published until 2018, that have covered test automation from requirements. According to the data provided in this SLR, only 21\% of the studies cover test automation from NL requirements, e.g.,~\cite{tahat2001requirement,ibrahim2007automatic,carvalho2014nat2testscr,zhang2014systematic,Wang:15}. This is primarily because several issues in NL requirements, e.g., incompleteness and ambiguity, make it challenging to automate automated test artefact generation~\cite{wang2020automatic,mustafa2021automated}. Hence, most of these existing approaches use templated requirements or some form of formalism~\cite{wang2020automatic,carvalho2014nat2testscr,nebut2006automatic}. Furthermore, test scenarios that are used in several organizations, including Austrian Post, are only seldom the format of test artefacts automatically generated from requirements. Very few research works have addressed this problem. For instance, Sarmiento et al.\cite{sarmiento2016test} utilize Petri-Nets to transform NL requirements into test scenarios. Cunning and Rozenbiit~\cite{cunning1999test} proposed an approach for generating sequenced scenarios with data parameters from the structured requirements specification. 


Our work in this paper leverages advanced NLP techniques, i.e., LLMs and retrieval-based augmentation to directly synthesize \emph{TSs} from NL requirements without the need for intermediate representations or formalizations. This approach allows for greater flexibility and adaptability to varied and complex requirement documents, addressing the challenge of domain-specific knowledge not extensively covered in public databases. We further address the \emph{TS} generation from bilingual requirements, which to the best of our knowledge, has not been addressed previously.


\begin{figure*}[!t]
  \includegraphics[width=0.85\textwidth]{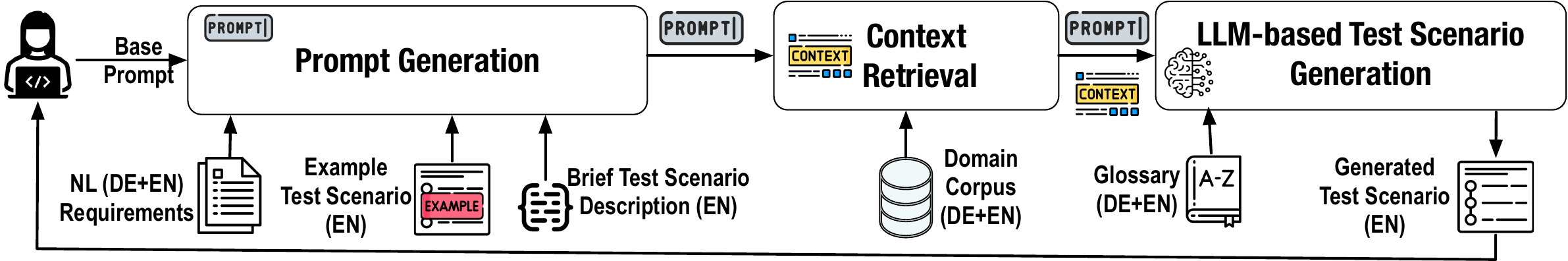}
  \centering
  \vspace*{-.7em}
  \caption{Retrieval Augmented Generation (RAG)-based Test Scenario Generation Approach.}
  \vspace*{-.5em}
  \label{fig:approach}
  \vspace*{-1em}
\end{figure*}

\section{Approach}~\label{sec:approach}
Fig.~\ref{fig:approach} provides an overview of our approach for Retrieval Augmented Generation-based Test scenArio Generation (\approach) from NL requirements. \approach~takes as input a set of NL requirements, an NL statement with a brief description of the test scenario to be generated, an example test scenario from the system (optional), a domain-information corpus, and a set of glossary terms (optional)~\cite{Arora:14a}. As mentioned in Section~\ref{subsec:Industry_Context}, in the context of this project with \"{O}PAG, the input NL requirements, glossary terms and domain-information corpus are written in a mix of German and English languages, while \emph{TS}-related information is in English. 

\subsection{Prompt Generation}
\begin{figure}  \includegraphics[width=0.45\textwidth]{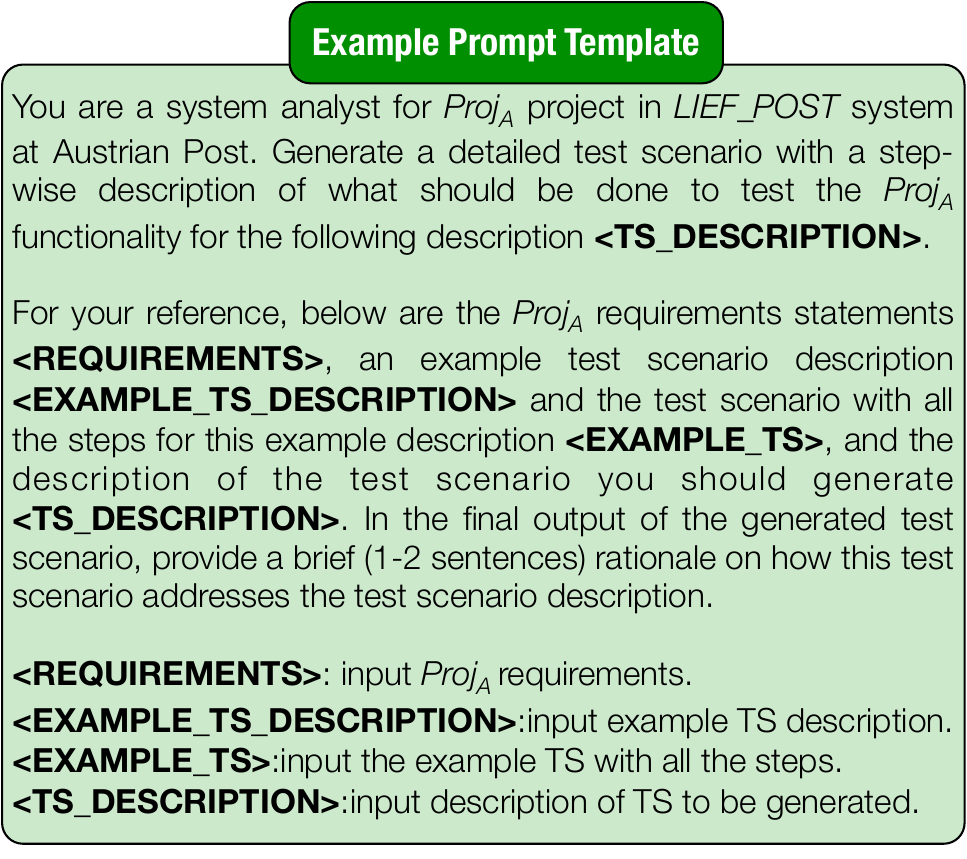}
  \centering
  \caption{Prompt Template used in \approach.}  
  \label{fig:promtTemplate}
\end{figure}

This step takes as input a set of NL requirements, an example \emph{TS} and a brief description of the \emph{TS} that needs to be generated. Providing the example \emph{TS} is optional. However, as we argue in Section~\ref{sec:evaluation}, providing the example yields better results.
The output of the step is a prompt which is passed onto the next steps.
This step assumes a pre-sectioned set of requirements based on features. In our projects, the requirements sets were already pre-sectioned. In the case of a new project with no sections, one can easily combine the requirements based on features or using NLP similarity techniques~\cite{ferrari2013using,ladeinde2023extracting}. The example \emph{TS} is required for few-shot prompting (discussed in Section~\ref{sec:background}) and guiding the LLM in generating \emph{TSs} in a desired format. The presence (or absence) of an example \emph{TS} is part of the subject of our analysis in RQ1. A brief description of the TS to be generated is required to explicitly tell the LLM about the coverage of specific context or conditions in the generated TS (e.g., ``Delivery with R\"{u}cksendung Ausland'' in Fig.~\ref{fig:example}). While LLMs also allow the generation of TSs without any description, such generated TSs can be very generic. In our preliminary experiments with Austrian Post Group IT (authors), we further observed the varying efficacy of LLMs in generating \emph{TSs} based on the richness and specificity of the input. This has implications for the extent of manual intervention required in the final \emph{TS} generation. Hence, \approach~requires a brief yet specific description of the TS to be generated. Fig.~\ref{fig:promtTemplate} shows the final prompt template used in \approach. This template results from numerous internal experimentation and fine-tuning iterations among the first two authors. 

\subsection{Context Retrieval}
Once the prompt is generated, the next step is to retrieve the related context from the RAG pipeline (Fig.~\ref{fig:RAGModel}). The indexing step of RAG takes the domain documentation corpus as input with HTML, PDF, Microsoft document, or text files and splits it into passages (or chunks). The default chunk size is 512 tokens, and we use the default size in our implementation as this is typically large enough to provide context. 
If the domain documentation is already indexed, this step simply fetches the context vectors from the vector DB (see Fig.~\ref{fig:RAGModel}).
As a part of this step, the input prompt from the previous step is transformed into the prompt embedding and passed onto the next step, along with the top-k context spans closely matching the prompt (querying stage of Fig.~\ref{fig:RAGModel}). 

\subsection{LLM-based Test Scenario Generation}
In this step, the TS is generated using an LLM based on the input prompt and the selected context passage(s) from the previous step. This step takes an optional input of a set of glossary terms (with definitions). As mentioned earlier, the glossary terms are required to provide additional context and maintain the specific keywords (e.g., German keywords in Fig.~\ref{fig:example} requirements) in the generated \emph{TS}. 
A subset of relevant glossary terms for a given project can be appended directly to the input prompt in this step. We do not append the glossary terms in the previous steps to avoid misleading the context retrieval.


Most LLMs have a token limit, e.g., the limit for GPT-3.5 turbo is 4,096 tokens. In case the combination of the input prompt, the selected context passage(s) and the glossary terms exceed this limit, certain strategies can be applied to fit the most context possible while complying with the token limit: (i) reduce the number of the retrieved context passages; and (ii) limit the requirements covered in the input prompt by selecting only a subset of relevant requirements. In our evaluation, we did not exceed this token limit even when prompting with top-3 context passages, all requirements and an example \emph{TS}.





\section{Evaluation}~\label{sec:evaluation}
This section presents our empirical evaluation settings and industrial projects details.

\subsection{Research Questions}
\sectopic{RQ1. Which \approach~configuration yields the most accurate results for \emph{TS} generation?} \approach~can be instantiated in numerous configuration settings, e.g., a different LLM, different prompting strategies, or retrieving more context chunks in RAG. By comparing alternatives, we plan to identify an accurate configuration for \approach. 

\sectopic{RQ2. Do practitioners find the \emph{TSs} generated by \approach~useful?} Ultimately, our \emph{TS} generation approach is valuable only if practitioners find the output of \approach~meaningful in their context. RQ2 aims to assess the perceptions of experts from our industry partner about the usefulness of the generated \emph{TSs}, using the selected configuration from RQ1.


\subsection{Implementation Details}~\label{subsec:ImplementationDetails}
\textcolor{black}{We have implemented \approach~in Python using the Azure OpenAI Services. We compare the OpenAI's GPT-3.5 (\textsf{gpt-3.5-turbo}) and GPT-4.0 (\textsf{gpt-4}) models in our evaluation at \textit{zero} temperature settings, i.e., setting the randomness component of the LLMs to zero.
We used Azure Cognitive Search for data indexing and retrieval. Using Azure OpenAI Studio, one can implement different strategies for retrieving the context chunks from the domain documentation and querying the LLMs. We use the `default' setting and explicitly set the maximum number of context spans ($k$). Chunking allows limiting the information sent to the LLMs due to token limits.}


\subsection{Project Details}~\label{subsec:ProjectDetails}
We perform our evaluation on two projects in \appName - \projOne~and \projTwo. \projOne~deals with all the processes associated with the deliveries to the special value-added-tax regions for our industry partner. \projTwo~deals with the processes related to the automated pre-sorting of all the posts before delivery for an efficient delivery process. Our selection of these two projects was driven by these criteria:
\sectopic{1. Access to at least one project with pre-specified requirements and corresponding \emph{TSs}.} RQ1 analysis required ground truth for determining the most effective configuration for \emph{TS} generation. Therefore, we sought projects that already had all this information available, as writing new \emph{TSs} for ground truth was beyond the scope of this study due to extensive effort and time commitment requirements from the experts.
\sectopic{2. Securing an adequate level of participation from the experts.} RQ2 analysis required the participation of experts in a data collection session. For this, we required projects the experts were reasonably familiar with, i.e., projects finished in the recent past or the ones in their late stages of development.
\sectopic{3. Project size should be commensurate with the usual projects at our industry partner.} We were interested in projects of reasonable size compared to other projects at Austrian Post Group IT. We also avoided targeting large projects in \appName. The rationale is that in RQ2 analysis, we wanted to elicit experts' feedback. Hence, to avoid overwhelming experts and having to rely on their memory of all the features and respect their time commitments, we only focused on projects of reasonable size in \appName. 

We selected \projOne~and \projTwo, as \projOne~met all the above criteria and \projTwo~met all the criteria, except the first one partially. We answer RQ1 using \projOne~data, and RQ2 using both the projects.

\subsection{Data Collection Procedure}~\label{subsec:dataCollection}
Our data collection was conducted over two stages. In the \emph{first stage}, we collected for both projects all the requirements, \emph{TS} descriptions, detailed \emph{TSs}, and the domain documentation. The authors first discussed the project selection criteria. The collaborators (second and third authors) collected all the requisite pre-existing data internally, vetted the data for any confidential information and completed two test descriptions missing in \projOne. There were a few other issues identified in the requirements statements that made them unconducive to automated analysis, e.g., convoluted structure or missing information as they implicitly referred to previously specified requirements. We deliberately decided not to address these issues, to be able to maintain the practical settings for \approach~evaluation. For domain documentation (Fig.~\ref{fig:approach}), the collaborators provided a manual that is used by \appName~end users and introduces most of the processes related to \appName. \textcolor{black}{We note that none of the \emph{TSs} were exposed to \approach, as part of the manual.} 
We note that all the requirements, \emph{TS} descriptions, and domain documentation collected in this stage were written in a mix of German and English. All the \emph{TSs} were in English, with domain-or system-specific keywords in German. Most glossary terms were available to us as part of the domain documentation. We extended this glossary with the specific keywords of \projOne~and \projTwo~missing from the domain documentation, with the help of our collaborators.
From \projOne, we collected 75 requirements statements (including three feature descriptions) and 16 detailed \emph{TSs} with descriptions. \projTwo~had 41 requirements statements (including two feature descriptions) and 15 test descriptions (with only partially written \emph{TSs} which were not considered in our analysis in RQ1 or RQ2).

In the \emph{second stage}, we collected the data from four experts in an interview survey to assess their perception of the quality of the generated \emph{TSs} for RQ2 analysis.  During this interview survey, the experts reviewed the \emph{TSs} from \projOne~and \projTwo~based on the five criteria listed below.

\begin{enumerate}[leftmargin=*]
    \item \textbf{Relevance} assesses how well the generated \emph{TS} aligns with the project requirements. The motivation for this criterion is to ensure that the generated TS directly contributes to validating the critical functionalities and objectives outlined in the requirements. 
    \item \textbf{Coverage} evaluates the `completeness' of the generated \emph{TS}, i.e., the extent to which the \emph{TS} encompasses all relevant aspects of the requirements and the test description. 
    \item \textbf{Correctness} evaluates the accuracy of \emph{TS} and its steps. It ensures the scenario is logically sound and covers the \emph{TS} steps in the correct sequence.
    \item \textbf{Understandability} assesses whether the generated \emph{TS} is clear and comprehensible to the experts, without any redundancies, so they can easily interpret and implement the \emph{TS} without ambiguity or confusion.  
    \item \textbf{Feasibility} evaluates whether the \emph{TS} can be executed with the available resources in the project setup. The motivation is to ensure that the \emph{TSs} are practical and can be realistically implemented within the project's constraints.
\end{enumerate}

Each criterion above was assessed by experts on a five-point Likert scale~\cite{likert1932technique} for the degree to which the generated \emph{TS} meets the criterion. \\
\emph{1. Strongly Disagree}: Significantly fails to meet the criterion. \\
\emph{2. Disagree}: Does not adequately meet the criterion. \\
\emph{3. Neutral}: Somewhat meets the criterion, but improvements are necessary.\\
\emph{4. Agree}: Meets the criterion satisfactorily, with only minor issues or omissions.\\
\emph{5. Strongly Agree}: Excellently meets the criterion.

We shared the information from both projects with the experts 48 hours before the planned session. 
At the start of the interview survey, we explained all the criteria to the experts and provided a quick recollection of the project requirements and features. We also used one generated \emph{TS} from \projOne~as an example discussion to explain the evaluation process. We, therefore, performed the actual evaluation on 15 \emph{TSs} each from \projOne~and \projTwo. For each criterion, the experts were asked to rate the generated \emph{TS} according to the five criteria based on the Likert scale and also verbalize their rationale for a given choice to ensure that they had a consistent understanding of each rating scale. The experts were free to provide individual ratings in the interview survey. However, in most cases, they all agreed on the same rating. In case, they gave a rating of 1, i.e., \emph{Strongly Disagree} on Relevance, Coverage or Correctness for any \emph{TS}, we did not proceed with the remaining criteria. Subsequently, we asked the experts to answer Feasibility based on the premise that the issues have been addressed. We also asked each expert to briefly share their overall perspective on \approach~at the end of the session. We recorded (wrote down) all their explanations in the session. 

\sectopic{Survey Participants.} The interview survey session had seven participants in total, including the researcher (first author), the facilitators (second and third authors) and four experts (respondents). The researcher led the survey and introduced the research project to the experts, and each generated \emph{TS}. The second author -- a scrum master at \company~ --  helped moderate the session and record the responses. The third author -- a requirements engineer at \company -- helped provide additional requirements and domain context, wherever necessary. The four experts were a quality assurance (testing) manager (pseudonym Gary), a product owner (pseudonym Rory), and two test engineers from \company~(pseudonyms Peter and Patty). Gary, Peter, and Patty have each worked at Austrian Post for three years.  Rory has been with Austrian Post for 33 years.
All four experts were involved in \projOne~and evaluated the generated \emph{TSs} using the aforementioned criteria. Three of the four experts were involved in \projTwo~and evaluated the corresponding \emph{TSs}. Peter was not part of \projTwo~and hence did not respond to any questions related to the project. The entire session lasted for a little over three hours for both projects, with two breaks of ten minutes each to mitigate fatigue. We report the results from the interview survey in RQ2 analysis (Section~\ref{subsec:discussion}).


\subsection{Evaluation Settings}~\label{subsec:EvaluationSettings}
We report on our evaluation settings for RQ1. We note that RQ2 was based on our interview survey, and all the relevant details have been discussed above.

\sectopic{RQ1 Evaluation.} For answering RQ1, in the prompt generation step of \approach, we experiment with two different settings -- zero-shot (ZS) and few-shot (FS) prompting. For zero-shot prompting, we exclude the example \emph{TS} (in Fig.~\ref{fig:promtTemplate}), its description or any reference to the example TS from the prompt sent to the retrieval and the test case generation stages. In the retrieval stage, we maintain the default configuration of the RAG model provided by the LlamaIndex library in Python, i.e., we index passages (or chunks) with a maximum limit of 512 tokens each. For the querying stage, we experiment with two settings for $k$, i.e., the number of context spans we experiment with $k=1$ and $k=3$. For the TS generation step of \approach~we experiment with two LLMs, i.e., GPT-3.5 and GPT-4.0. Hence, in total, we experiment with eight configurations in RQ1. 

\textit{Metrics.} In the NLP domain, various metrics are used to evaluate the quality of text generation. Three widely recognized metrics are BLEU~\cite{papineni2002bleu}, ROUGE~\cite{lin2004rouge}, and METEOR~\cite{banerjee2005meteor}. BLEU measures the correspondence between a machine-generated output and a human by comparing the co-occurrence of n-grams (word pairs, triples, etc.) in the generated text to a set of reference texts and calculates a score based on this comparison. High BLEU scores indicate a greater similarity to the reference texts, suggesting better output quality. ROUGE is typically used in summarization tasks. It assesses the quality of a generated text by measuring how many words and phrases it has in common with a reference text, focusing on recall. METEOR, another metric for evaluating translation quality, extends metrics such as BLEU beyond simple n-gram matching. It incorporates synonyms and stemming, aligning more closely with the human judgment of generated text quality. METEOR compares the generated text to reference texts, accounting for word order. It produces a score based on precision and recall, offering a balanced view of translation performance. We use all three metrics (normalized on a scale of 0-1) to evaluate the quality of generated \emph{TSs} in RQ1.

\subsection{Results and Discussion}\label{subsec:discussion}
%
\begin{table}
\centering
\caption{RQ1 Results.}
\label{tab:RQ1_Results}
\vspace*{-1em}
\begin{threeparttable}[t]
\begin{tabularx}{0.49\textwidth}{@{} p{0.08\textwidth} @{\hskip 0.5em}  p{0.05\textwidth} p{0.05\textwidth} @{\hskip 0.5em} *{3}{>{\centering\arraybackslash}X}@{}}
    \toprule
LLM &  Prompt & Chunks &  BLEU & ROUGE & METEOR \\ 
\midrule
GPT3.5 & ZS  & 1 & 0.004 & 0.151  & 0.229   \\   
GPT3.5 & ZS  & 3 & 0.005 & 0.161  & 0.238  \\   
GPT3.5 & FS   & 1 & {0.084} & \textbf{0.419} & \textbf{0.528}   \\ 
GPT3.5 & FS   & 3 & {0.080} & {0.406} & {0.470}   \\   
GPT4.0 & ZS   &1 &  {0.003} & {0.149}  & 0.244   \\
GPT4.0 & ZS   &3 &  {0.005} & {0.159}  & 0.236   \\
GPT4.0 & FS   &1 &  \textbf{0.092} & \textbf{0.419}  & 0.516   \\
GPT4.0 & FS   &3 &  {0.067} & 0.373  & 0.467   \\
\bottomrule
\end{tabularx}
 \end{threeparttable}
 \vspace*{-2em}
\end{table}
\sectopic{RQ1.} Table~\ref{tab:RQ1_Results} shows the BLEU, ROUGE, and METEOR scores for different combinations experimented in RQ1 (Section~\ref{subsec:EvaluationSettings}). 
For both GPT-3.5 and GPT-4.0, few-shot (FS) prompting yields significantly higher BLEU, ROUGE, and METEOR scores than zero-shot (ZS), indicating that providing examples improves accuracy and relevance in generated TSs. Hence, as widely known in the NLP community, FS prompting can lead to better results (than ZS prompting), particularly when the possibility of building a fine-tuned LLM via labelled data is expensive~\cite{perez2021true}.

Regarding the number of chunks considered for the RAG model, increasing the number of context chunks from 1 to 3 does not consistently improve performance metrics across both models and prompting strategies, implying that adding context must be balanced and relevant. A common trend visible from Table~\ref{tab:RQ1_Results} is that increasing the number of context chunks leads to improved results for ZS prompting but not for FS prompting. One plausible explanation for this could be that in the absence of much context in ZS prompting, additional context helps. In contrast, for FS prompting, where examples already provide specific guidance, additional context (from different parts of the domain documentation) might be unnecessary and introduce confusion if it diverges from the generation task's core requirements.

For the comparison between the two GPT models considered in our evaluation, GPT-4.0 with FS prompting achieves the highest BLEU score, indicating better exact match quality. However, GPT-4.0's ROUGE and METEOR scores in some setups are comparable or slightly lower than GPT-3.5's, suggesting nuances in how each model handles NL understanding and generation. We also note a qualitative finding from reviewing all the TSs generated, that most GPT-4.0 TSs are more verbose than those generated by GPT-3.5. Based on these observations, we select the configuration of GPT-3.5 with $k=1$ and FS prompting for answering RQ2. The absolute metrics for this configuration suggest that while the exact phrasing of the generated \emph{TSs} may not closely match the actual \emph{TSs} (BLEU=0.092), the essential content (ROUGE=0.419) and overall meaning (METEOR=0.528) are reasonably well captured.

\vspace{.1em}
\begin{tcolorbox}[arc=0mm,width=\columnwidth,
                  top=0mm,left=0mm,  right=0mm, bottom=0mm,
                  boxrule=1pt]
\textit{The answer to RQ1 is that the GPT-3.5 LLM with few-shot prompting and the top retrieved-context passage in the retrieval stage of \approach~leads to the best results for test scenario generation.}
\end{tcolorbox}

\begin{figure}[!t]
\includegraphics[width=0.39\textwidth]{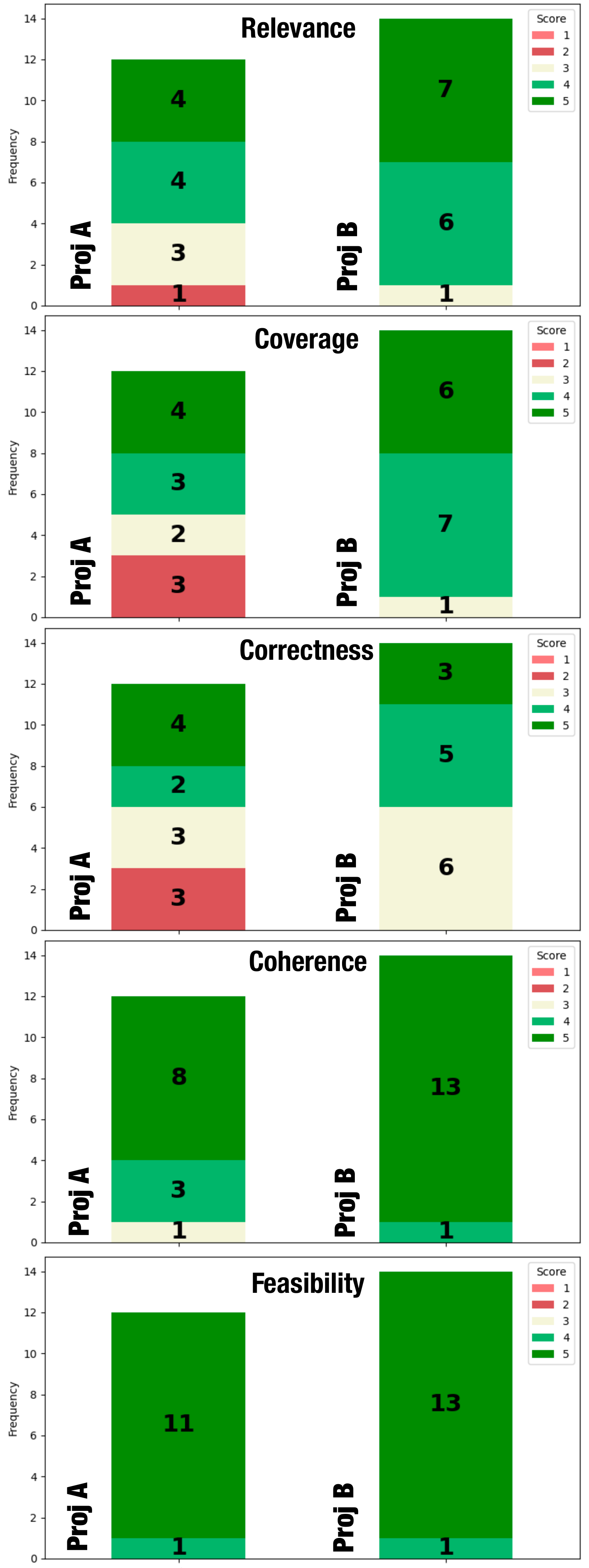}
  \centering
  \vspace*{-.7em}
  \caption{RQ2 Results with the frequency of Likert Scale ratings for each evaluation criterion by experts.}
  \vspace*{-.5em}
  \label{fig:RQ2_Analysis}
  \vspace*{-1em}
\end{figure}

\sectopic{RQ2.} Fig.~\ref{fig:RQ2_Analysis} shows the overall results from our expert interview survey. 
Overall, in 11 out of the 30 (36.7\%) \emph{TSs}, the experts (strongly) agreed on all criteria, i.e., the generated scenarios were of very high quality. In 11 out of the 30 (36.7\%) \emph{TSs} the experts (strongly) agreed on most criteria and were neutral on one or two criteria. For 4 of the 30 (13.3\%) scenarios, the experts disagreed on at least one of the five criteria. For the remaining four \emph{TSs}, the experts strongly disagreed on the test scenario's relevance, coverage or correctness. We did not proceed with a rating for understandability or feasibility in these cases. Of these four cases, two test scenarios generated by LLMs were deemed completely wrong, i.e., the test scenarios had major issues, particularly with the correctness of the steps in the test scenarios in \projOne. For example, one test scenario generated misleading information, as the steps were not in line with the actual steps in \appName. For the remaining two test scenarios in this category (one each from \projOne~and\projTwo), the experts mentioned that there were obvious quality issues in the requirements and/or the test description, and minor amendments could have potentially led to better test scenarios. However, we did not have an opportunity to make any improvements or perform an iterative analysis in our evaluation. \textcolor{black}{Hence, we marked all criteria for these test scenarios as rating 1 (\emph{Strongly Disagree}). We exclude these scenarios from all subsequent analyses to avoid bias due to the low-quality inputs.}


The overall results of 26 test scenarios in two projects, with averages ($\mu$) of assessment criteria with standard deviation ($\sigma$), are as follows:
Relevance $\mu (\sigma)$ = 4.19 (0.85),
Coverage $\mu (\sigma)$ = 4.04 (1.0),
Correctness $\mu (\sigma)$ = 3.69 (1.0),
Coherence $\mu (\sigma)$ = 4.77 (0.51),
Feasibility $\mu (\sigma)$ = 4.92 (0.27).

\textit{Relevance.} $\mu = 4.19$ suggests that our experts agree that the automatically generated \emph{TSs} are closely aligned with the requirements. $\sigma = 0.85$ indicates some variability in experts' perception of relevance. However, the variation is relatively moderate, suggesting reasonably consistent perceptions. One \emph{TS} in \projOne~did not adequately cover the concepts in the requirements and the test description (i.e., received a rating of \emph{Disagree} or 2 from the experts). The experts were \emph{Neutral} on four \emph{TSs}, as the generated \emph{TS} did not fully align with the intended description. In two of these instances, Rory noted issues in the descriptions, which require expertise to fully comprehend the intended direction for generating the \emph{TSs}.

\textit{Coverage.} $\mu = 4.04$ suggests that experts agree that the \emph{TSs} cover the most relevant concepts from the requirements and the test description, with minor omissions. $\sigma=1.0$ denotes noticeable variability, indicating that some test scenarios were perceived as more comprehensive than others. From Fig.~\ref{fig:RQ2_Analysis}, five \emph{TSs} in \projOne~did not satisfactorily cover the concepts in the requirements and the test description, as these \emph{TSs} missed the specific steps required to reach a particular state in \appName.

\textit{Correctness.} $\mu = 3.69$ implies moderate accuracy in \emph{TSs} or more than expected omissions or issues.
$\sigma=1.0$ indicates significant variability. This suggests that evaluators' perceptions of correctness varied more widely than other aspects, pointing to some inconsistencies or fluctuating levels of accuracy in the generated \emph{TSs}. One of the key issues in several \emph{TSs} with rating 1--3 for Correctness was missing or inaccurate steps (related to domain and project understanding of \appName) from the \emph{TS}.

\textit{Coherence.} A high $\mu = 4.77$   indicates that the \emph{TSs} were generally clear and easy to comprehend.
The relatively low $\sigma=0.51$ reflects a consistent level of understandability across test scenarios, suggesting that they followed consistent domain terminology.

\textit{Feasibility.} The highest $\mu=4.92$ implies that the \emph{TSs} are viewed as highly practical and executable (post addressing the coverage or correctness issues).
The $\sigma=0.27$ highlights a strong consensus among experts regarding the feasibility of \emph{TS} execution, suggesting that the LLMs did not introduce any infeasible steps or hyperbolic conditions, ensuring that the generated \emph{TSs} remain grounded in practicality.

\textcolor{black}{After the interview survey, we also reviewed all the expert feedback. We note that overall generated \emph{TSs} in \projOne~had a lower rating than \projTwo~in general (see Fig.~\ref{fig:RQ2_Analysis}). For example, three of the four \emph{TSs}, where experts strongly disagreed on relevance, coverage or correctness, were in \projOne. A key difference between the two projects is that, due to the nature of \projOne~(VAT calculation), it had stricter requirements for following the sequence of certain steps and phases in \appName~than \projTwo. The kind of mistakes made by \approach~were similar across some generated \emph{TSs}, which could have been potentially addressed via (minor) adjustments to the prompts outside evaluation settings. However, as noted earlier, we did not allow the experiments' prompts and outputs or iterations to be fixed to avoid bias or influence our evaluation results.}

\vspace{.1em}
\begin{tcolorbox}[arc=0mm,width=\columnwidth,
                  top=0mm,left=0mm,  right=0mm, bottom=0mm,
                  boxrule=1pt]
\textit{The answer to RQ2 is that the practitioners strongly agree that the \emph{TSs} generated by \approach~are coherent and feasible. The relevance and coverage are also deemed satisfactory, with only minor adjustments, indicating the meaningfulness of the generated \emph{TSs} and alignments to the requirements. Correctness shows high variability, suggesting areas for improvement regarding accuracy and adherence to the specific \emph{TS} steps and a need for more specific domain knowledge in \approach.}
\end{tcolorbox}

\subsection{Expert Feedback Discussion}~\label{sec:implications}
\approach~has been subjected to expert scrutiny in our interview survey. During the survey, the experts verbalized their rationale for any omissions/inaccuracies/issues in the generated \emph{TS}, and also provided their overall perception of \approach, at the end. In this section, we provide a brief qualitative analysis of their feedback and the implications of this research for Austrian Post Group IT.

\sectopic{Insights on Translation and Domain-Specific Knowledge.}
Experts highlighted the challenge of maintaining domain-specific terms. Gary suggested, \textit{``Maybe I would recommend making a dictionary of terms that should not be translated other way than specified, and have that dictionary as a source for an LLM,''} pointing towards a more nuanced handling of bilingual requirements. While we had attempted this by using the glossary terms, GPT models inexplicably ignored glossary terms in some \emph{TSs} and translated them to English, impacting correctness. 

\sectopic{Quality of Input Data and Reverse Engineering.}
The quality of input data was underscored as a pivotal factor in the performance of RAG/LLMs by all experts and the two industry authors, also a known issue in RAG~\cite{barnett2024seven}. Rory noted, \textit{``if LLM gets input data which is low quality or too ambiguous, it will also have such output."} 
This emphasizes the need for clear and precise requirements documentation and the \emph{TS} description.
An indirect advantage of \approach~noted by experts is that in case the generated TS has major issues, it most likely is due to poor quality of input requirements. Hence, they can reflect on the original requirements and locate issues that led to incorrectness. We posit that this reverse engineering exercise might help improve the quality of original requirements; however, an in-depth investigation is required to establish this.

\sectopic{Incorporating Broader System Information.}
The potential for LLMs to enhance \emph{TS} generation through access to system architecture was discussed. Rory suggested, \textit{``Based on my observations, it [the approach] can perform better in creation of \emph{TS} if it would have access to the architecture schema of the used systems/applications,"} advocating for a more integrated approach to scenario generation.

\sectopic{Experience and Human Oversight Matters.}
The value of human expertise in refining LLM-generated \emph{TSs} was acknowledged. Patty mentioned \textit{``More experienced testers understand or refine the LLM generated scenarios much easier,''}. Peter further said \textit{``It [The approach] is missing feedback loop. Some scenarios can be easily fixed.''} highlighting the indispensable role of human oversight and experience.


\sectopic{Feedback and Future Directions.}
All experts agreed that they would use \approach~to automate \emph{TS} generation in their projects. Patty noted that \textit{``It seems like a useful tool, and much better than writing test scenarios manually. They require minor adjustments but that is much easier than writing it from scratch.''} Gary further noted that \textit{``As we noted, some of the scenarios were completely off. Such cases are inevitable but are very easy to write off by just looking at them. Hence, the technology is worth it.''}
The concept of using LLMs to map related test scenarios was proposed as an innovative direction with the potential to streamline test scenario management. Additionally, the positive reception from Peter—\textit{``Can we use this for our actual project next week?"}—signals a readiness to integrate RAGTAG into real-world projects, contingent upon further refinement and validation.

In conclusion, the feedback from experts provides a clear directive towards enhancing \approach~with specific focus on improving translation accuracy, input data quality, system architecture integration, and leveraging experienced tester insights for refinement.

\vspace*{-.3em}
\section{Threats to Validity}~\label{sec:threats}
\vspace*{-1.1em}
\sectopic{Internal Validity.} Bias is one of the main concerns for internal validity. To mitigate bias, the experts in RQ2 had no exposure to our technical approach. Further, there was a possibility of social desirability bias, where experts might refrain from expressing dissenting opinions—we implemented several strategies to counteract this. We provided opportunities for each expert to voice their opinions, asked them to verbalize their rationales, and randomly alternated the sequence in which they reviewed the \emph{TSs}. The experts did note their disagreement in some cases (Section~\ref{subsec:discussion}).

\sectopic{External Validity.} Generalizability is always a concern with an industry-innovation study like ours. We evaluated \approach~on two different projects, but more diverse case studies at Austrian Post and beyond are required to ensure that \approach~is effective at generating \emph{TSs} from NL requirements. \textcolor{black}{We further plan to conduct more user studies, wherein the experts can execute the generated \emph{TSs} in practice to substantiate the results reflected by our evaluation criteria.}



\section{Conclusion}~\label{sec:conclusion}
In this paper, we presented an automated approach (\approach) for generating test scenarios from NL requirements, developed in collaboration with Austrian Post Group IT -- a division of Austrian Post. \approach~is based on large language models (LLMs) and retrieval augmented generation (RAG) techniques. We reported on the evaluation of \approach~on two projects from Austrian Post, as a part of the \appName~software system used for postage and delivery management with bilingual requirements, specified in German and English. In this context, we evaluated eight configurations for \approach~and identified the best configuration, with GPT-3.5 LLM and few-shot prompting. The paper further reports on an interview survey conducted with four experts from Austrian Post on the usefulness of \approach~in practice, in terms of relevance, coverage, correctness, coherence and feasibility of the generated test scenarios. The results show that experts (strongly) agree that the generated test scenarios are coherent, feasible, relevant and cover the relevant concepts. They also indicated the need for a human expert in the loop to improve correctness. \approach~according to the experts is ready for adoption in their context and is likely to save time in their quality assurance efforts. In future, we would like to conduct wider studies at Austrian Post and beyond to establish the generalizability of \approach, and extend the domain documentation to include the architectural information of the systems. \textcolor{black}{We also plan to attempt wider experiments with more LLMs and RAG parameter settings, e.g., temperature in LLMs.}


\sectopic{Acknowledgement. }
We gratefully acknowledge the efforts invested by the colleagues at Austrian Post, in providing feedback at different stages in this project.


\bibliographystyle{IEEEtran}
\balance

\bibliography{paper}

\begin{thebibliography}{10}
\providecommand{\url}[1]{#1}
\csname url@samestyle\endcsname
\providecommand{\newblock}{\relax}
\providecommand{\bibinfo}[2]{#2}
\providecommand{\BIBentrySTDinterwordspacing}{\spaceskip=0pt\relax}
\providecommand{\BIBentryALTinterwordstretchfactor}{4}
\providecommand{\BIBentryALTinterwordspacing}{\spaceskip=\fontdimen2\font plus
\BIBentryALTinterwordstretchfactor\fontdimen3\font minus
  \fontdimen4\font\relax}
\providecommand{\BIBforeignlanguage}[2]{{%
\expandafter\ifx\csname l@#1\endcsname\relax
\typeout{** WARNING: IEEEtran.bst: No hyphenation pattern has been}%
\typeout{** loaded for the language `#1'. Using the pattern for}%
\typeout{** the default language instead.}%
\else
\language=\csname l@#1\endcsname
\fi
#2}}
\providecommand{\BIBdecl}{\relax}
\BIBdecl

\bibitem{desikan2006software}
S.~Desikan and G.~Ramesh, \emph{Software testing: principles and
  practice}.\hskip 1em plus 0.5em minus 0.4em\relax Pearson Education India,
  2006.

\bibitem{crispin2009agile}
L.~Crispin and J.~Gregory, \emph{Agile testing: A practical guide for testers
  and agile teams}.\hskip 1em plus 0.5em minus 0.4em\relax Pearson Education,
  2009.

\bibitem{unterkalmsteiner2014taxonomy}
M.~Unterkalmsteiner, R.~Feldt, and T.~Gorschek, ``A taxonomy for requirements
  engineering and software test alignment,'' \emph{ACM Transactions on Software
  Engineering and Methodology (TOSEM)}, vol.~23, no.~2, pp. 1--38, 2014.

\bibitem{gordon2009generating}
M.~Gordon and D.~Harel, ``Generating executable scenarios from natural
  language,'' in \emph{International Conference on Intelligent Text Processing
  and Computational Linguistics}.\hskip 1em plus 0.5em minus 0.4em\relax
  Springer, 2009, pp. 456--467.

\bibitem{nayak2011synthesis}
A.~Nayak and D.~Samanta, ``Synthesis of test scenarios using uml activity
  diagrams,'' \emph{Software \& Systems Modeling}, vol.~10, no.~1, pp. 63--89,
  2011.

\bibitem{cunning1999test}
S.~J. Cunning and J.~Rozenbiit, ``Test scenario generation from a structured
  requirements specification,'' in \emph{Proceedings ECBS'99. IEEE Conference
  and Workshop on Engineering of Computer-Based Systems}.\hskip 1em plus 0.5em
  minus 0.4em\relax IEEE, 1999, pp. 166--172.

\bibitem{klischat2019generating}
M.~Klischat and M.~Althoff, ``Generating critical test scenarios for automated
  vehicles with evolutionary algorithms,'' in \emph{2019 IEEE Intelligent
  Vehicles Symposium (IV)}.\hskip 1em plus 0.5em minus 0.4em\relax IEEE, 2019,
  pp. 2352--2358.

\bibitem{bertolino2006product}
A.~Bertolino, A.~Fantechi, S.~Gnesi, and G.~Lami, ``Product line use cases:
  Scenario-based specification and testing of requirements,'' in \emph{Software
  Product Lines}.\hskip 1em plus 0.5em minus 0.4em\relax Springer, 2006, pp.
  425--445.

\bibitem{wang2020automatic}
C.~Wang, F.~Pastore, A.~Goknil, and L.~C. Briand, ``Automatic generation of
  acceptance test cases from use case specifications: an nlp-based approach,''
  \emph{IEEE Transactions on Software Engineering}, vol.~48, no.~2, pp.
  585--616, 2020.

\bibitem{hois2014natural}
B.~Hois, S.~Sobernig, and M.~Strembeck, ``Natural-language scenario
  descriptions for testing core language models of domain-specific languages,''
  in \emph{2014 2nd International Conference on Model-Driven Engineering and
  Software Development (MODELSWARD)}.\hskip 1em plus 0.5em minus 0.4em\relax
  IEEE, 2014, pp. 356--367.

\bibitem{cem2013introduction}
J.~Cem~Kaner, ``An introduction to scenario testing,'' \emph{Florida Institute
  of Technology, Melbourne}, pp. 1--13, 2013.

\bibitem{Arora:23}
C.~Arora, J.~Grundy, and M.~Abdelrazek, ``Advancing requirements engineering
  through generative ai: Assessing the role of llms,'' \emph{arXiv preprint
  arXiv:2310.13976}, 2023.

\bibitem{kandpal2023large}
N.~Kandpal, H.~Deng, A.~Roberts, E.~Wallace, and C.~Raffel, ``Large language
  models struggle to learn long-tail knowledge,'' in \emph{International
  Conference on Machine Learning}.\hskip 1em plus 0.5em minus 0.4em\relax PMLR,
  2023, pp. 15\,696--15\,707.

\bibitem{lewis2020retrieval}
P.~Lewis, E.~Perez, A.~Piktus, F.~Petroni, V.~Karpukhin, N.~Goyal,
  H.~K{\"u}ttler, M.~Lewis, W.-t. Yih, T.~Rockt{\"a}schel \emph{et~al.},
  ``Retrieval-augmented generation for knowledge-intensive nlp tasks,''
  \emph{Advances in Neural Information Processing Systems}, vol.~33, pp.
  9459--9474, 2020.

\bibitem{gao2023retrieval}
Y.~Gao, Y.~Xiong, X.~Gao, K.~Jia, J.~Pan, Y.~Bi, Y.~Dai, J.~Sun, and H.~Wang,
  ``Retrieval-augmented generation for large language models: A survey,''
  \emph{arXiv preprint arXiv:2312.10997}, 2023.

\bibitem{sabetzadeh2024practical}
M.~Sabetzadeh and C.~Arora, ``Practical guidelines for the selection and
  evaluation of {NLP} techniques in {RE},'' \emph{arXiv preprint
  arXiv:2401.01508}, 2024.

\bibitem{nguyen2023generative}
A.~Nguyen-Duc, B.~Cabrero-Daniel, A.~Przybylek, C.~Arora, D.~Khanna, T.~Herda,
  U.~Rafiq, J.~Melegati, E.~Guerra, K.-K. Kemell \emph{et~al.}, ``Generative
  artificial intelligence for software engineering--a research agenda,''
  \emph{arXiv preprint arXiv:2310.18648}, 2023.

\bibitem{Jurafsky:20}
D.~Jurafsky and J.~H. Martin, \emph{Speech and Language Processing}, 3rd~ed.,
  2020, \url{https://web.stanford.edu/~jurafsky/slp3/}(visited 2021-06-04).

\bibitem{brown2020language}
T.~B. Brown, B.~Mann, N.~Ryder, M.~Subbiah, J.~Kaplan, P.~Dhariwal,
  A.~Neelakantan, P.~Shyam, G.~Sastry, A.~Askell, S.~Agarwal, A.~Herbert-Voss,
  G.~Krueger, T.~Henighan, R.~Child, A.~Ramesh, D.~M. Ziegler, J.~Wu,
  C.~Winter, C.~Hesse, M.~Chen, E.~Sigler, M.~Litwin, S.~Gray, B.~Chess,
  J.~Clark, C.~Berner, S.~McCandlish, A.~Radford, I.~Sutskever, and D.~Amodei,
  ``Language models are few-shot learners,'' 2020.

\bibitem{reimers2019sentence}
N.~Reimers and I.~Gurevych, ``Sentence-bert: Sentence embeddings using siamese
  bert-networks,'' \emph{arXiv preprint arXiv:1908.10084}, 2019.

\bibitem{manning1999foundations}
C.~Manning and H.~Schutze, \emph{Foundations of statistical natural language
  processing}, 1999.

\bibitem{mustafa2021automated}
A.~Mustafa, W.~M. Wan-Kadir, N.~Ibrahim, M.~A. Shah, M.~Younas, A.~Khan,
  M.~Zareei, and F.~Alanazi, ``Automated test case generation from
  requirements: A systematic literature review,'' \emph{Computers, Materials
  and Continua}, vol.~67, no.~2, pp. 1819--1833, 2021.

\bibitem{tahat2001requirement}
L.~H. Tahat, B.~Vaysburg, B.~Korel, and A.~J. Bader, ``Requirement-based
  automated black-box test generation,'' in \emph{25th annual international
  computer software and applications conference. COMPSAC 2001}.\hskip 1em plus
  0.5em minus 0.4em\relax IEEE, 2001, pp. 489--495.

\bibitem{ibrahim2007automatic}
R.~Ibrahim, M.~Z. Saringat, N.~Ibrahim, and N.~Ismail, ``An automatic tool for
  generating test cases from the system's requirements,'' in \emph{7th IEEE
  International Conference on Computer and Information Technology (CIT
  2007)}.\hskip 1em plus 0.5em minus 0.4em\relax IEEE, 2007, pp. 861--866.

\bibitem{carvalho2014nat2testscr}
G.~Carvalho, D.~Falcao, F.~Barros, A.~Sampaio, A.~Mota, L.~Motta, and
  M.~Blackburn, ``Nat2testscr: Test case generation from natural language
  requirements based on scr specifications,'' \emph{Science of Computer
  Programming}, vol.~95, pp. 275--297, 2014.

\bibitem{zhang2014systematic}
M.~Zhang, T.~Yue, S.~Ali, H.~Zhang, and J.~Wu, ``A systematic approach to
  automatically derive test cases from use cases specified in restricted
  natural languages,'' in \emph{System Analysis and Modeling: Models and
  Reusability: 8th International Conference, SAM 2014, Valencia, Spain,
  September 29-30, 2014. Proceedings 8}.\hskip 1em plus 0.5em minus 0.4em\relax
  Springer, 2014, pp. 142--157.

\bibitem{Wang:15}
C.~Wang, F.~Pastore, A.~Goknil, L.~Briand, and Z.~Iqbal, ``Automatic generation
  of system test cases from use case specifications,'' in \emph{Proceedings of
  the 2015 International Symposium on Software Testing and Analysis - ISSTA
  2015}, 2015.

\bibitem{nebut2006automatic}
C.~Nebut, F.~Fleurey, Y.~Le~Traon, and J.-M. Jezequel, ``Automatic test
  generation: A use case driven approach,'' \emph{IEEE Transactions on Software
  Engineering}, vol.~32, no.~3, pp. 140--155, 2006.

\bibitem{sarmiento2016test}
E.~Sarmiento, J.~C. Leite, E.~Almentero, and G.~S. Alzamora, ``Test scenario
  generation from natural language requirements descriptions based on
  petri-nets,'' \emph{Electronic Notes in Theoretical Computer Science}, vol.
  329, pp. 123--148, 2016.

\bibitem{Arora:14a}
C.~Arora, M.~Sabetzadeh, L.~Briand, and F.~Zimmer, ``Improving requirements
  glossary construction via clustering: approach and industrial case studies,''
  in \emph{Proceedings of the 8th ACM/IEEE International Symposium on Empirical
  Software Engineering and Measurement (ESEM'14)}, 2014.

\bibitem{ferrari2013using}
A.~Ferrari, S.~Gnesi, and G.~Tolomei, ``Using clustering to improve the
  structure of natural language requirements documents,'' in \emph{Requirements
  Engineering: Foundation for Software Quality: 19th International Working
  Conference, REFSQ 2013, Essen, Germany, April 8-11, 2013. Proceedings
  19}.\hskip 1em plus 0.5em minus 0.4em\relax Springer, 2013, pp. 34--49.

\bibitem{ladeinde2023extracting}
A.~Ladeinde, C.~Arora, H.~Khalajzadeh, T.~Kanij, and J.~Grundy, ``Extracting
  queryable knowledge graphs from user stories: An empirical evaluation,'' in
  \emph{International Conference on Evaluation of Novel Approaches to Software
  Engineering 2023}.\hskip 1em plus 0.5em minus 0.4em\relax Scitepress, 2023,
  pp. 684--692.

\bibitem{likert1932technique}
R.~Likert, ``A technique for the measurement of attitudes.'' \emph{Archives of
  psychology}, 1932.

\bibitem{papineni2002bleu}
K.~Papineni, S.~Roukos, T.~Ward, and W.-J. Zhu, ``{Bleu}: a method for
  automatic evaluation of machine translation,'' in \emph{Proceedings of the
  40th annual meeting of the Association for Computational Linguistics}, 2002,
  pp. 311--318.

\bibitem{lin2004rouge}
C.-Y. Lin, ``{Rouge}: A package for automatic evaluation of summaries,'' in
  \emph{Text summarization branches out}, 2004, pp. 74--81.

\bibitem{banerjee2005meteor}
S.~Banerjee and A.~Lavie, ``{METEOR}: An automatic metric for mt evaluation
  with improved correlation with human judgments,'' in \emph{Proceedings of the
  acl workshop on intrinsic and extrinsic evaluation measures for machine
  translation and/or summarization}, 2005, pp. 65--72.

\bibitem{perez2021true}
E.~Perez, D.~Kiela, and K.~Cho, ``True few-shot learning with language
  models,'' \emph{Advances in neural information processing systems}, vol.~34,
  pp. 11\,054--11\,070, 2021.

\bibitem{barnett2024seven}
S.~Barnett, S.~Kurniawan, S.~Thudumu, Z.~Brannelly, and M.~Abdelrazek, ``Seven
  failure points when engineering a retrieval augmented generation system,''
  \emph{arXiv preprint arXiv:2401.05856}, 2024.

\end{thebibliography}

\end{document}